\documentclass[prc,preprint,showpacs,showkeys,tightenlines]{revtex4}

\usepackage{graphicx}  %
\usepackage{bm}  %

\newcommand{\beq}{\begin{equation}}
\newcommand{\eeq}{\end{equation}}
\newcommand{\bea}{\begin{eqnarray}}
\newcommand{\eea}{\end{eqnarray}}

\newcommand{\kf}{k_{\rm F}}

\newcommand{\order}[1]{{\cal O}(#1)}          

\def\vec#1{{\bf #1}}    

\newcommand{\psihat}{\widehat\psi}
\newcommand{\xvec}{\vec{x}}
\newcommand{\dagphan}{{\phantom{\dagger}}}
\newcommand{\kvec}{\vec{k}}

\newcommand{\ak}{a^\dagphan_\kvec}
\newcommand{\akdag}{a^\dagger_\kvec}

\newcommand{\nab}{\overrightarrow{\nabla}}
\newcommand{\nabsq}{\overrightarrow{\nabla}^{2}\!}

\newcommand{\galnab}{\tensor{\nabla}}
\newcommand{\psid}{{\psi^\dagger}}
\newcommand{\psidagger}{{\psi^\dagger}}
\newcommand{\idt}{{i\partial_t}}

\newcommand{\Left}{{\cal L}}
\newcommand{\Lefta}{{\cal L}_\alpha}

\begin{document}

\title{Are Occupation Numbers Observable?}

\author{R. J. Furnstahl}\email{furnstahl.1@osu.edu} 
\author{H.-W. Hammer}\email{hammer@mps.ohio-state.edu}

\affiliation{Department of Physics,
         The Ohio State University, Columbus, OH\ 43210, USA}

%
\date{August 28, 2001}

\begin{abstract}
The question of whether occupation numbers and momentum
distributions of nucleons in nuclei are observables is
considered from an effective field theory perspective.
Field redefinitions lead to variations that imply the answer is
negative, as illustrated in the interacting Fermi gas at low density.
Implications for the interpretation of
$(e,e'p)$ experiments with nuclei are discussed.
\end{abstract}

\smallskip
\maketitle

The advent of new continuous beam electron accelerators (CEBAF, Mainz,
MIT-Bates, NIKHEF) permits experiments that probe hadronic matter in 
both inclusive and exclusive reactions with unprecedented precision. 
These experiments are expected to deepen our understanding of nuclear 
structure and reaction mechanisms by measuring the response of
nuclei to electroweak probes over a wide range of energy and momentum 
transfers  \cite{BOFFI96,PANDHARIPANDE97}. 
Exclusive electron 
scattering experiments at large momentum transfer, such as the knockout
of a proton in $(e,e'p)$ on a nucleus, are particularly important. 

It is often claimed that occupation numbers and/or momentum distributions 
of nucleons in nuclei can be extracted from these experiments. 
Their extraction from the data is built on 
the extreme impulse approximation but is obscured by the need to consider 
final state interactions and meson exchange currents.
The interpretation and comparison with theory is made based on a given 
nuclear Hamiltonian, but from the perspective of the underlying theory of 
the strong interaction, QCD, there is no unique or preferred Hamiltonian. 
Rather, there are infinitely many such low-energy effective
Hamiltonians that are related by field redefinitions that leave 
observables unchanged (e.g., see Ref.~\cite{LEPAGE89}).
What happens to occupation numbers under such transformations?

A powerful framework to study low-energy phenomena
in a model-independent way is given by effective field theory 
(EFT) \cite{LEPAGE89,Birareview,BEANE99}.
The underlying idea is to exploit a separation of scales in the system.
For example, if the typical momenta $k$ are small compared to the 
inverse range of the interaction $1/R$, 
low-energy observables can be described by a
controlled expansion in $kR$. All short-distance effects are systematically
absorbed into low-energy constants using renormalization.
The EFT approach allows for accurate calculations of low-energy processes 
and properties with well-defined error estimates.

In this letter, we explore from an EFT perspective the question of
whether occupation numbers and momentum
distributions of nucleons in nuclei are observable.
In an EFT, observables are characterized by invariance under local field 
redefinitions. If a quantity depends on the particular representation of 
the Lagrangian ${\cal L}$ (beyond the level of truncation errors), 
it is not an observable.
Off-shell Green's functions for scattering processes in the vacuum, 
for example, can be changed by field redefinitions. On-shell
Green's functions, which correspond to S-matrix elements, however, 
are unchanged \cite{HAAG58,COLEMAN69}. 

To address the issue most cleanly, we focus on the question of
whether occupation numbers in a homogenous medium
at finite density are observables 
in the framework of EFT. 
In a general finite system, occupation numbers and momentum distributions 
are very different quantities. 
In a homogeneous system, however, they
are equivalent.
Since there are no asymptotic states in an infinite medium, however, 
the usual analysis for field redefinitions 
does not directly carry over to in-medium observables. 
While the analysis can be extended to thermodynamic observables
like the energy density or particle number \cite{FURNSTAHL01},
this extension is not obvious for other quantities, such as the
momentum distribution or occupation numbers.

We use the  interacting Fermi gas 
at low density as a laboratory to illustrate a
fundamental and generic problem with the definition of 
momentum occupation numbers.
For a given representation of the Hamiltonian, one can define
an operator $\akdag \ak$ that counts particles/holes with 
momentum $\kvec$ and gives the expected result in the noninteracting limit.
As discussed below, this operator 
is not derived from a global symmetry,
in contrast to the total number operator. 
In the EFT framework
this is a problem, as there is no preferred form of the 
effective Lagrangian. 

To highlight this problem, we consider both the particle number $N$
and the momentum distribution $n(p)$.
We describe the system using
a local Lagrangian for a nonrelativistic fermion
field with spin independent interactions
that is invariant under Galilean, parity, and time-reversal
transformations:
\bea
  {\cal L}  &=&
       \psi^\dagger \biggl[i\partial_t + \frac{\nabsq}{2M}\biggr]
                 \psi - \frac{C_0}{2}(\psi^\dagger \psi)^2
            + \frac{C_2}{16}\Bigl[ (\psi\psi)^\dagger
                                  (\psi\galnab^2\psi)+\mbox{ H.c.}
                             \Bigr]  +  \ldots
\,,\label{lag}
\eea
where $\galnab=\overleftarrow{\nabla}-\nab$ is the Galilean invariant
derivative and H.c.\ denotes the Hermitian conjugate.
A convenient and transparent regularization scheme is dimensional
regularization with minimal subtraction; see, e.g., Ref.~\cite{HAMMER00} 
for details. In this scheme the coefficients are simply
$C_0 = 4\pi a/M$ and $C_2=C_0 a r_e/2$ where $a$ is 
the  $s$-wave scattering length and $r_e$ the effective
range.

We generate an infinite, equivalent class of Lagrangians 
$\Lefta$ by performing the field redefinition
\bea
\psi &\to& \psi+\frac{4\pi\alpha}{\Lambda^3}(\psi^\dagger \psi)\psi\,,
    \qquad
    \psi^\dagger \to \psi^\dagger+\frac{4\pi\alpha}{\Lambda^3}\psi^\dagger
    (\psi^\dagger \psi)\,,
       \label{ftrafo}
\eea
in ${\cal L}$. The factor $1/\Lambda^3$ is introduced  to keep
the arbitrary parameter $\alpha$ dimensionless. $\Lambda$ is the 
breakdown scale of the EFT and the additional factor
of $4\pi$ is introduced for convenience \cite{FURNSTAHL01}. 
We obtain for $\Lefta$:
\bea
\Lefta &=&{\cal L}-\frac{4\pi\alpha}{\Lambda^3}2 C_0 (\psid\psi)^3
      \nonumber\\
 &+&\frac{4\pi\alpha}{\Lambda^3}\left\{(\psid\psi)\psid(\idt\psi)
   -\frac{1}{2M}\left[\psid(\nab\psi)\cdot\psid(\nab\psi)+2(\nab
   \psid)\psi\cdot\psid(\nab\psi)\right]+\mbox{ H.c.} \right\}
      \nonumber\\
 &+& \cdots + {\cal O}(\alpha^2)\,,
\label{lagtrans}
\eea
where higher-order two- and three-body terms, 
all four- and higher-body terms, and terms of ${\cal O}(\alpha^2)$
have been omitted.  

For $\alpha=0$ we recover the original $\Left$.
The Lagrangian $\Lefta$ contains additional vertices, including 
an off-shell vertex, but gives exactly the same energy density
and particle number as the Lagrangian ${\cal L}$.  
In Ref.~\cite{FURNSTAHL01} it was illustrated how the necessary
cancellations occur in general and for this particular example.
Furthermore, it was shown how the number operator
must be constructed as the conserved charge 
of the Noether current associated with the $U(1)$ phase symmetry:
\beq
   \psi(x) \longrightarrow e^{-i\phi}\psi(x) \quad \mbox{ and } \quad
   \psid(x) \longrightarrow e^{i\phi}\psid(x)\ ,
\eeq
under which $\Lefta$ is invariant.
One can conveniently identify the Noether current by promoting $\phi$
to a function of $x$ and considering infinitesimal
transformations with
\beq
  {\cal L}_\alpha \longrightarrow \widetilde{\cal L}_\alpha
   [\psi,\psidagger;\phi(x)]
   \ .
\eeq
Then the number density operator is given by  \cite{FURNSTAHL01}:
\beq
   \label{defNalpha}
   \widehat{N}^\alpha  \equiv  \frac{\delta}{\delta(\partial_t \phi)}
           \widetilde{\cal L}_\alpha [\psi,\psidagger;\phi(x)]
           = \psid\psi + \frac{4\pi\alpha}{\Lambda^3}
        2 \left(\psid\psi \right)^2 \ .
\eeq
The particle number itself is given by the spatial integral of
$\widehat{N}^\alpha$. In a uniform system, however, the difference is
simply a factor of the volume and it is convenient to 
refer to $\widehat N^\alpha$ as the number operator. Note that
Eq.~(\ref{defNalpha}) differs from the naive expectation $\psid\psi$ 
for $\alpha \neq 0$.
In the appendix of Ref.~\cite{FURNSTAHL01}, it was demonstrated how
the contributions from the additional vertices in $\Lefta$ 
and the additional term in $ \widehat{N}^\alpha$ cancel order-by-order 
in $\alpha$. Consequently, the total particle number $N$ is unchanged by 
field redefinitions, as expected for an observable.

\begin{figure}[t]
\begin{center}
\includegraphics[width=3in,angle=0,clip=true]{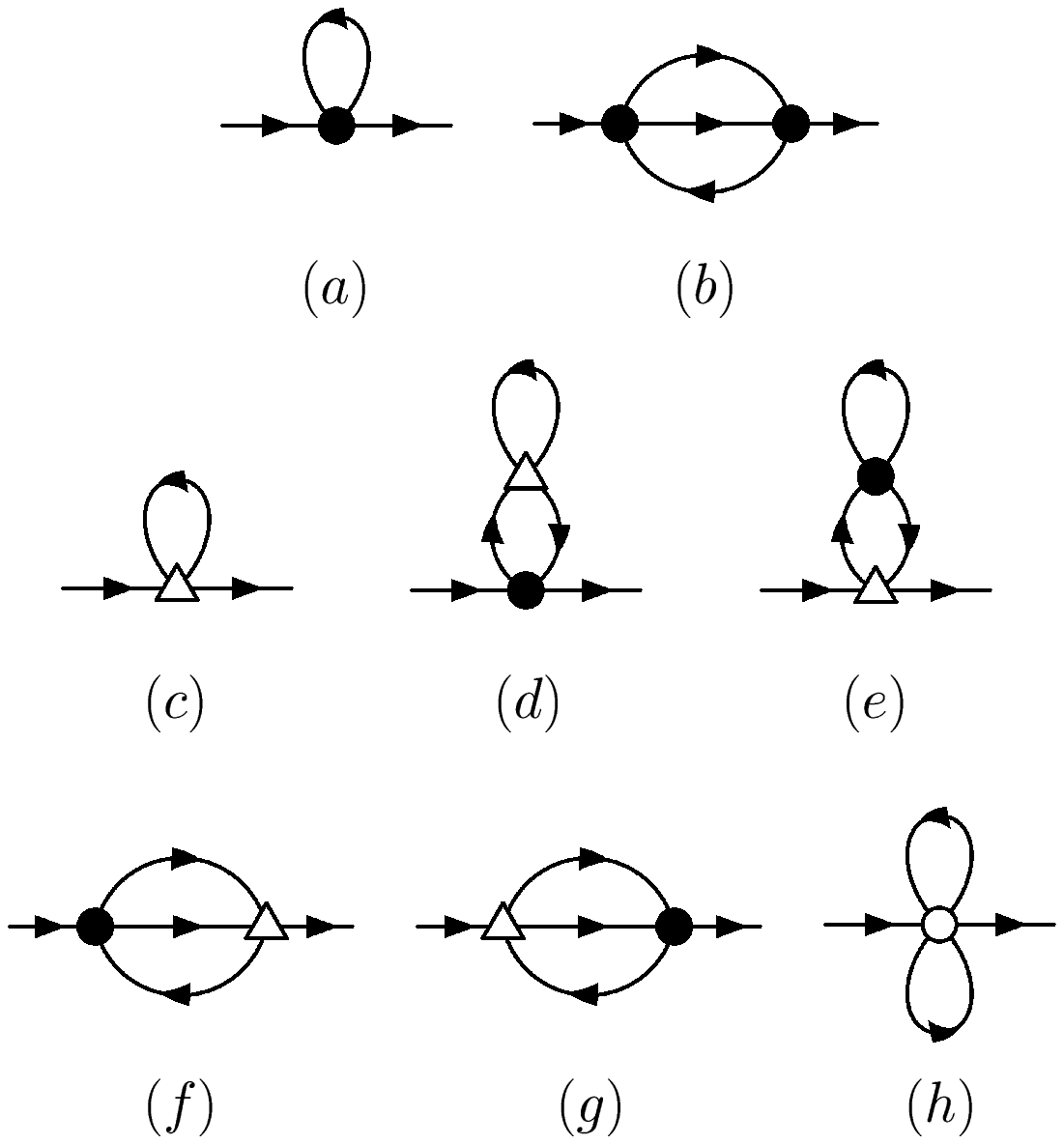}
\end{center}
\vspace*{-18pt}
\caption{Diagrams contributing to $\Sigma^*(\omega,\kvec)$ 
to $\order{C_0^2}$ [(a) and (b)] and at $\order{\alpha C_0}$
[(c) to (h)].}
\label{fig:self}
\end{figure}

Matters become more complicated for the momentum distribution
$n(k)$ since the corresponding operator is not simply related to the
conserved charge of a Noether current. In the literature, the operator
\beq 
\label{npnaiv}
\widehat{n}_\kvec=\akdag \ak \,,
\label{npop}
\eeq
where $\ak$ destroys a fermion of momentum ${\bf k}$,
is universally adopted \cite{BOFFI96,PANDHARIPANDE97,BRANDOW67,MIGDAL67}.
This definition gives the correct 
result in the noninteracting limit, $n(k)=g\, \theta(k_F -k)$,
where $g$ is the spin degeneracy factor. If there were a 
preferred form of the Lagrangian/Hamiltonian, as is usually assumed, 
Eq.~(\ref{npop}) would uniquely determine the momentum
distribution. In terms of the field operators
\beq
  \label{eq:fexp}
  \psihat(\xvec) = \int\! \frac{d^3 k}{(2\pi)^3}\, e^{i\kvec\cdot\xvec} \ak \ ,
  \qquad
  \psihat^\dagger(\xvec) =\int\! \frac{d^3 k}{(2\pi)^3}\,
      e^{-i\kvec\cdot\xvec} \akdag\ ,
\eeq        
we can write $\akdag \ak = \int d^3 x\, \widehat{n}_\kvec(\xvec)$ with
\beq
\widehat{n}_\kvec(\xvec)=\int d^3 y\; e^{i\kvec\cdot\vec{y}}
            \psihat^\dagger(\xvec+\vec{y})\psihat(\xvec)\,,
\eeq
which is nonlocal in coordinate space.
Using the definition of the one-particle Green's function \cite{FETTER71},
\beq
iG(\xvec,t;\xvec',t')=\frac{\langle \Psi_0 | T [ \psihat_H (\xvec,t)
\psihat^\dagger_H(\xvec',t')] | \Psi_0\rangle}{\langle \Psi_0 |\Psi_0
\rangle}\,,
\eeq
with $ |\Psi_0\rangle$ the exact ground state of the system and 
\beq
\psihat_H (\xvec,t) = e^{i\widehat{H} t} \psihat (\xvec)
e^{-i\widehat{H} t}\,,
\eeq
a Heisenberg operator, the expectation value of
$\widehat{n}_\kvec(\xvec)$ can be written as \cite{MIGDAL67}
\beq
\label{defocc}
n(k) = \langle \widehat{n}_\kvec(\xvec) \rangle= \lim_{\eta\to 0^+}
(-i) g \int \frac{d\omega}
{2\pi} e^{i\omega\eta} G(\omega, \kvec)\,.
\eeq
The momentum distribution $n(k)$ for the hard sphere Fermi gas 
to second order in $\kf a$ has been calculated using this definition 
in Refs.~\cite{BELYAKOV61,SARTOR80}.

%
\begin{figure}[t]
\begin{center}
\includegraphics[width=3.0in,angle=0,clip=true]{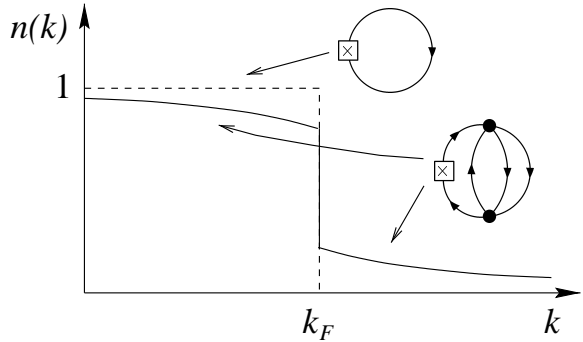}
\end{center}
\vspace*{-18pt}
\caption{Schematic picture of the occupation number as a function
of momentum in a uniform Fermi system with no interactions
(dashed line) and including leading correction from interactions
(solid line) (cf. Ref.~\protect\cite{BRANDOW67}). The square
with a cross denotes an insertion of Eq.~(\ref{ins_npnaiv}).
}
\label{fig:occupation}
\end{figure}
%

With the definition in Eq.~(\ref{defocc}), the explicit expressions
for $n(k)$ in Refs.~\cite{BELYAKOV61,SARTOR80} 
are reproduced in
the EFT by setting $\alpha=0$ and
calculating the one-particle Green's function to
$\order{C_0^2}$. The one-particle Green's function $G(\omega,\kvec)$
is related to the proper self energy $\Sigma^*(\omega,\kvec)$ via
\cite{FETTER71}
\beq
G(\omega,\kvec)=\frac{1}{\omega-\kvec^2/(2m)-\Sigma^*(\omega,\kvec)}\,,
\eeq
where the spin indices have been suppressed. To $\order{C_0^2}$
there are only two diagrams for the proper self energy, which
are shown in Fig.~\ref{fig:self}(a) and (b). The filled
circle represents the $C_0$ interaction from Eq.~(\ref{lag}). 
The Feynman rules for evaluating these diagrams can be found 
in Refs.~\cite{HAMMER00,FURNSTAHL01}.
Figure~\ref{fig:occupation} shows schematically how the
$\order{C_0^2}$ contribution modifies the distribution.
Since particles can be kicked out of the Fermi sea by the interaction, 
some occupation
probability is moved to states above the Fermi surface.
(Note that the second-order diagram shown is a Feynman diagram and
so includes modifications of both particle and hole states.)

\begin{figure}[t]
\begin{center}
\includegraphics[width=4.0in,angle=0,clip=true]{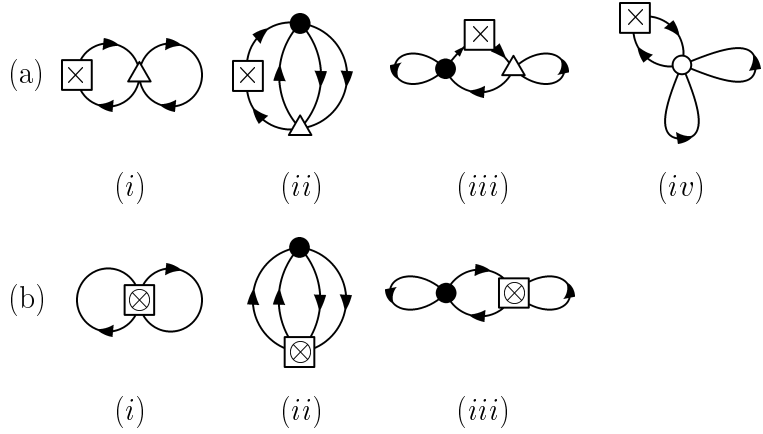}
\end{center}
\vspace*{-18pt}
\caption{Feynman diagrams for the occupation number $n(k)$.
(a) gives the contribution from $\protect\akdag\protect\ak$ while (b) shows
contributions from the additional term for $\alpha\neq 0$.
Note that diagram (a)$(iii)$ represents only one of three possible
insertions of $\protect\akdag\protect\ak$. The square with a cross and
the square with a circled cross denote insertions of 
Eqs.~(\ref{ins_npnaiv}) and (\ref{ins_np}), respectively.
}
\label{fig:Nkalpha}
\end{figure}

If we use the equivalent Lagrangian $\Lefta$
with $\alpha\neq0$, $n(k)$ differs already at $\order{\alpha}$. 
The additional self-energy diagrams up to $\order{\alpha C_0}$ are shown
in Fig.~\ref{fig:self} (c)--(h). Here the empty circle represents
the induced three-body vertex $\propto \alpha C_0$ and the empty
triangle the induced off-shell vertex proportional to $\alpha$
in Eq.~(\ref{lagtrans}). (The Feynman
rules for these vertices can again be found in 
Refs.~\cite{HAMMER00,FURNSTAHL01}.) 
Equivalently, the occupation numbers
can be calculated from energy diagrams with an
operator insertion corresponding to Eq.~(\ref{npnaiv}).
The appropriate insertion for $\akdag\ak$ on a fermion line with 
energy $\omega$ and momentum $\vec{p}$ is  
\beq
\label{ins_npnaiv}
(-i) (2\pi)^3 \lim_{\eta\to 0^+} e^{i\omega\eta }
\delta^3 (\vec{p}-\vec{k}) \delta_{\alpha\beta}\,,
\eeq
where $\alpha$ and $\beta$ are spin indices and the factor
$\exp(i\omega \eta)$ ensures the correct ordering of operators. 
The corresponding diagrams
up to $\order{\alpha C_0}$ are shown in Fig.~\ref{fig:Nkalpha}(a),
where we have indicated the insertion of Eq.~(\ref{ins_npnaiv})
by the square with the cross. (Note that the diagram in 
Fig.~\ref{fig:Nkalpha}(a)$(iii)$ represents only one of 
three possible insertions of the operator $\akdag\ak$.) 
Only the first two diagrams
in Fig.~\ref{fig:Nkalpha}(a) are nonvanishing. We find
\bea
\Delta n(k)_{(i)} &=& -2(g-1)\rho\frac{4\pi\alpha}{\Lambda^3}\theta(\kf-k)  
\nonumber \\
\Delta n(k)_{(ii)} &=& 2ig(g-1) \frac{4\pi\alpha}{\Lambda^3} \frac{C_0}
{(2\pi)^9} \lim_{\eta\to 0^+} \int d^4 p \int d^4 l \int d^4 q\, 
e^{i\omega\eta} \nonumber\\
& &\times \delta^3 (\vec{p}-\vec{k}) G_0(p) G_0(l) G_0(p-q) G_0(l+q)\,.
\label{modnk}
\eea
This discrepancy is not surprising, since the definition in 
Eq.~(\ref{defocc}) corresponds to the operator given in Eq.~(\ref{npnaiv}).
Even the operator for the total particle number corresponding to 
the Lagrangian $\Lefta$ is different from the naive expectation.
For the occupation number, however, there is no symmetry that
can be used to construct the operator $\widehat{n}^\alpha_\kvec$ 
and so an ambiguity occurs.
This simple exercise can be continued to higher order, with increasingly
sophisticated diagrams.
The same pattern recurs, with additional diagrams depending on $\alpha$
induced at each order, generating $\alpha$ dependence in $n(k)$.
{\it This implies that $n(k)$ is not an observable.}

We might take Eq.~(\ref{npnaiv}) together with ${\cal L}$ as a 
{\em definition\/}
and transform the operator $\widehat{n}_\kvec$ at the same time as ${\cal L}$.
For finite $\alpha$ this implies that one has to calculate the 
additional diagrams shown in Fig.~\ref{fig:Nkalpha}(b) where the
square with the circled cross denotes an insertion of
\beq
\label{ins_np}
(-i) (2\pi)^3 \frac{4\pi\alpha}{\Lambda^3} \left( \delta_{\alpha_1 \alpha_3}
\delta_{\alpha_2 \alpha_4} +  \delta_{\alpha_1 \alpha_4} 
\delta_{\alpha_2 \alpha_3} \right)
\lim_{\eta\to 0^+} \sum_{j=1}^4 e^{i\omega_{j} \eta} \delta^3 (\vec{p}_j
-\vec{k} )\,,
\eeq
where $\omega_j$, $\vec{p}_j$, and $\alpha_j$ label the 
energy, momentum, and spin of of the line $j$, respectively. 
The occupation numbers defined this way are independent of $\alpha$
by construction.
The first two diagrams in Fig.~\ref{fig:Nkalpha}(b)
exactly cancel the contributions from the first two diagrams in
Fig.~\ref{fig:Nkalpha}(a) (cf. Eq~(\ref{modnk})),
while the third diagram vanishes.
{\it However, we had no basis for the original definition, since
EFT is model independent and makes no a priori assumptions on the 
dynamics. } Thus we conclude that occupation numbers (or 
even momentum distributions) cannot be uniquely defined in general.

How does this conclusion fit in with the standard analysis of
$(e,e'p)$ experiments, where 
the cross sections measured are, by definition, observables?
There are further complications in a finite system, where
the momentum distribution differs from the occupation numbers,
but we can illustrate the analogous situation in our model problem
by introducing an external source
$J(x)$ coupled to the fermion number.
[Thus $J(x)$ plays the role of the Coulomb field of the virtual photon
in $(e,e'p)$.]
If we were constructing an EFT of an underlying theory (such as QCD), 
we would expect to need the most general coupling consistent
with the symmetries, but for simplicity we will assume that $J(x)$
has nonzero coupling only to $\psi^\dagger \psi$ 
in the original representation ${\cal L}$.

\begin{figure}[t]
\begin{center}
\includegraphics[width=3.5in,angle=0,clip=true]{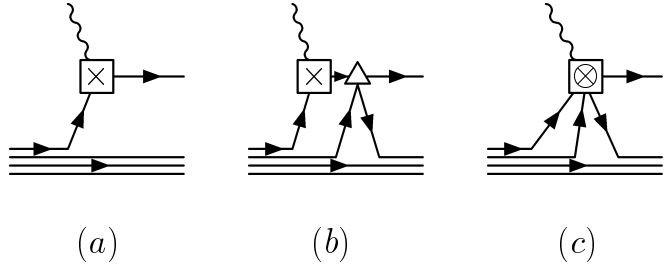}
\end{center}
\vspace*{-18pt}
\caption{Schematic diagrams for the interaction of an external
source $J(x)$ [wavy line] with the fermion system,
``knocking out'' a particle.}
\label{fig:knockout}
\end{figure}

The noninteracting cross section for $\alpha=0$ corresponds to the
diagram in Fig.~\ref{fig:knockout}(a).
The same cross section is obtained for $\alpha\neq 0$, but only if
we include the
contributions from the induced vertex to the final state interaction
in Fig.~\ref{fig:knockout}(b) [and to the initial state interaction,
which is not shown]
{\em and\/} the vertex contribution from the modified operator in 
Fig.~\ref{fig:knockout}(c).
In general, there are {\em always\/} contributions of all three types,
dressed with additional interactions order-by-order, which
are mixed up under field redefinitions.
Isolating Fig.~\ref{fig:knockout}(a) is a model-dependent procedure
since it depends on $\alpha$.

Note that the ambiguities have a natural size,
as discussed for an analogous shifting of contributions between two-body
off-shell and three-body vertices in Ref.~\cite{FURNSTAHL01}.
An interesting question is whether the stark difference in occupation
numbers between nonrelativistic and relativistic Brueckner calculations
\cite{JAMINON90} can be explained by the ambiguity.
Similar ambiguities occur in other areas of physics as well. In 
deep inelastic scattering, the physical cross section can be written
as a convolution of quark and gluon distributions with coefficient
functions determined by perturbative QCD. 
It is well known, however, that the distributions and
the coefficient functions are individually scheme and scale dependent.
The scheme and scale dependence of auxiliary quantities such as the pion 
distribution in a nucleon was recently clarified in Ref.~\cite{CHEN01}.
Another question of current interest is whether the condensate 
fraction in a Bose-Einstein condensate, which is essentially the 
occupation number of the condensate, can be measured \cite{LEGGETT01}.

The conclusion that the extraction of a momentum distribution
from $(e,e'p)$ cross sections is ambiguous because of final state
interactions and vertex corrections (e.g., meson exchange currents)
is not a surprise.
While it is well known that such ambiguities are present, the usual
assumption is  that there is a ``correct'' answer that can be
extracted from experiment. In a similar spirit, different ways of
implementing the impulse approximation for the response 
of many-fermion systems have been analyzed in Ref.~\cite{BENHAR01}.
In contrast,
the EFT perspective clearly implies that ambiguities  in the
extraction of momentum distributions 
in $(e,e' p)$  cannot be resolved by experiment. 
It is not only that the momentum distribution is difficult to extract 
but that it cannot be isolated in principle within a calculational
framework based on low-energy degrees of freedom.
Rather, such auxiliary quantities can only be defined in 
a specific convention, like a particular form of the Hamiltonian,
regularization scheme and so on.  It can still be useful 
to discuss such quantities within a given convention.
All true observables, however,  can just as well be described
in a different framework that adheres to different conventions.

\acknowledgments

We thank E.~Braaten, H.~Grie\ss hammer, S.~Jeschonnek, X.~Ji,
R.~Perry, S.~Puglia, and B.~Serot for useful comments.
This work was supported in part by the U.S. National Science
Foundation under Grant Nos.\ PHY--9800964 and PHY--0098645.

\end{document}